\documentclass[aps,pra,groupedaddress,showpacs]{revtex4}

\usepackage{epsfig}
\usepackage{graphicx}
\usepackage{amsmath}
\usepackage{amsfonts,amssymb}
\usepackage{color}
\usepackage{subfigure}

\begin{document}


\title{Dual Lindstedt series and Kolmogorov-Arnold-Moser theorem}


\author{Marco Frasca}
\email[e-mail:]{marcofrasca@mclink.it}
\affiliation{Via Erasmo Gattamelata, 3 \\
             00176 Roma (Italy)}


\date{\today}

\begin{abstract}
We prove that exists a Lindstedt series that holds when a Hamiltonian is driven by a perturbation going to infinity. This series appears to be dual to a standard Lindstedt series as it can be obtained by interchanging the role of the perturbation and the unperturbed system. The existence of this dual series implies that a dual KAM theorem holds and, when a leading order Hamiltonian exists that is non degenerate, the effect of tori reforming can be observed with a system passing from regular motion to fully developed chaos and back to regular motion with the reappearance of invariant tori. We apply these results
to a perturbed harmonic oscillator proving numerically the appearance of tori reforming. Tori reforming appears as an effect limiting chaotic behavior to a finite range of parameter space of some Hamiltonian systems.
Dual KAM theorem, as proved here, applies when the perturbation, combined with a kinetic term, provides again an integrable system.
\end{abstract}


\maketitle


\section{Introduction}

Perturbation theory is one of the oldest tools to solve differential equations. The idea behind is to identify a small parameter in the equation and then find a solution series with such a parameter. This approach can find its main limitation in the inherent difficulty in a lot of problems where such a small parameter appears difficult to identify or does not exist at all. In this case one has to cope with a non-perturbative problem with very few methods to work with other than numerical computations. This is the typical situation of non-linear dynamical systems that are quite generic in producing chaotic solutions. A chaotic regime arises when a given threshold is overcome and this happens when one or more parameters become large. In this situation there is no way to get analytical solutions unless the system is integrable.

The main aim of our analysis will be to manage a Hamiltonian system when a large perturbation is applied. What we are going to study is a dual situation with respect a weak perturbation. In this latter case the mathematical limit it taken with the parameter going to zero while in our case the mathematical limit corresponds on taking this same parameter going to infinity. We have set the stage in Ref.\cite{fra1} where the idea of a duality principle in perturbation theory was put forward and applied to quantum mechanics. Since then, we have extended this idea to several fields of physics ranging from quantum optics \cite{fra2,fra3,fra4} to general relativity \cite{fra5} and quantum field theory \cite{fra6,fra7,fra8}. Duality principle in perturbation theory can be stated by saying that interchanging perturbation terms into an equation produces two perturbation series having the development parameters one the inverse of the other. So, these two series apply in different regimes where the parameter goes to zero and goes to infinity. We have got a strong coupling expansion.

Hamiltonian systems are central in the description of most physical phenomena. Integrability is an exception and we can treat them, for most cases, only through perturbation theory. Indeed, the key approach for them is given by computing a
Lindstedt series \cite{arn0}.  
The convergence of this series has been a serious problem since Kolmogorov, Arnol'd and Moser proved a fundamental theorem \cite{kolm,arn2,arn3,mos1,mos2} that shows how singular terms, due to resonances, do not harm the behavior of the series. Indeed, invariant tori are shown to be preserved by a small perturbation.
When the perturbation increases, invariant tori become progressively destroyed until, for a given threshold, fully chaotic motion sets in. The aim of our paper is to see the situation the other way around. We assume a strong perturbation and analyze the dual Lindstedt series to see if KAM theorem can yet be applied. So, we will be able to formulate a dual KAM theorem. A dual KAM theorem implies that, for a very large perturbation, invariant tori can be seen to reappear turning back chaotic motion to a regular one.

KAM theorem has a large body of literature. Quite recent works apply renormalization group concepts to prove it \cite{gal1,bgk,eds} but our aims here are not to describe all this notable history but rather to try to see if there is another perspective to know the behavior of Hamiltonian systems in a different range of parameter space. The reason for this is that, being such systems ubiquitous, the possible applications could be a large number.

The idea is that a dual KAM theorem should hold when interchanging the choice of the perturbation does not change the quasi-periodical nature of the system. We expect a change in the resonances due to a different unperturbed evolution, but maintaning the same initial conditions. Anyhow, implied in this view is the idea that the choice of the unperturbed system does not make impossible to get the corresponding solution. 
Finally, dual KAM theorem is effective only when an interchanging of the perturbation terms yields again an integrable system for the leading order.

The paper has the following structure. In sec. \ref{ls} we give notations and present Lindstedt series. In sec. \ref{dp} we formulate the duality principle for Hamiltonian systems. In sec. \ref{kam} we formulate KAM theorem with a dual Lindstedt series. 
In sec. \ref{app} an applications to a well-known system is presented. Finally, in sec. \ref{cn} we give the conclusions.

\section{Lindstedt series\label{ls}}

We consider a Hamiltonian system having
\begin{equation}
    H=H_0({\bf p},{\bf q})+\lambda H_1({\bf p},{\bf q},t)
\end{equation}
so that
\begin{eqnarray}
    \dot q_i(t)&=&\frac{\partial H_0}{\partial p_i}+\lambda\frac{\partial H_1}{\partial p_i} \\ \nonumber
    \dot p_i(t)&=&-\frac{\partial H_0}{\partial q_i}-\lambda\frac{\partial H_1}{\partial q_i}.
\end{eqnarray}
When $\lambda\ll 1$ we are supposed to be able to find a solution in the form of a Lindstedt series
\begin{eqnarray}
    q_i(t)&=&\sum_{n=0}^\infty \lambda^nq_i^n(t) \\ \nonumber
    p_i(t)&=&\sum_{n=0}^\infty \lambda^np_i^n(t)
\end{eqnarray}
when these series exist and converge and has no secular terms. The leading order solution is given by the equations
\begin{eqnarray}
    \dot q_i^0(t)&=&\frac{\partial H_0}{\partial p^0_i} \\ \nonumber
    \dot p_i^0(t)&=&-\frac{\partial H_0}{\partial q^0_i}
\end{eqnarray}
and we are supposed to know their solution. This means that $H_0$ belongs to the class of integrable systems. Indeed, for most applications, we are interested to small perturbations to known systems that we are able to manage.


On the same ground, we can define a dual Lindstedt series as
\begin{eqnarray}
\label{eq:dlin}
    q_i(t)&=&\lambda^\alpha\sum_{n=0}^\infty \lambda^{-n}\hat q_i^n(t) \\ \nonumber
    p_i(t)&=&\lambda^\beta\sum_{n=0}^\infty \lambda^{-n}\hat p_i^n(t)
\end{eqnarray}
that holds in the limit $\lambda\rightarrow\infty$ and with $\alpha$ and $\beta$ to be fixed. Our aim will be to find out the leading order equations with the higher order corrections. 

\section{Duality principle for Hamiltonian systems\label{dp}}

Duality principle in perturbation theory can be stated in the following way \cite{fra1}:

{\bf Definition:} {\sl For a differential equation defined through a differential operator $L$ acting on a function $u$ in such a way to have
\begin{equation}
   Lu+V_0(u,u',u'',\ldots)+V_1(u,u',u'',\ldots)=0
\end{equation}
the solution series obtained taking $V_1$ as a perturbation is dual the one obtained taking $V_0$ as a perturbation, the development parameters being one the inverse of the other.
}

Now we use this definition showing the existence of a dual Lindstedt series for Hamiltonian systems. Let us consider the following Hamiltonian
\begin{equation}
   H=\frac{{\bf p}^2}{2m}+V_0({\bf p},{\bf q})+\lambda V_1({\bf p},{\bf q})
\end{equation}
where we have omitted explicit time dependence for the sake of simplicity. In the limit $\lambda\rightarrow 0$ we will have that the unperturbed system will be driven by
\begin{equation}
   H_0 = \frac{{\bf p}^2}{2m}+V_0({\bf p},{\bf q})
\end{equation}
as this is the Hamiltonian obtained in the given limit. In the opposite limit $\lambda\rightarrow\infty$, the question to be answered is if the momenta are bounded in this case. They are not. So, let us rescale $p_i\rightarrow\sqrt{\lambda}p_i$, we will get
\begin{equation}
   H=\lambda\frac{{\bf p}^2}{2m}+V_0({\bf p},{\bf q})+\lambda V_1({\bf p},{\bf q})
\end{equation}
and finally, rescaling $H\rightarrow\lambda H$, we are able to obtain
\begin{equation}
\label{eq:res}
   H'=\frac{{\bf p}^2}{2m}+V_1({\bf p},{\bf q})+\frac{1}{\lambda} V_0({\bf p},{\bf q})
\end{equation}
and we recognize that the perturbation parameter is now $1/\lambda$ and the role of perturbed and unperturbed terms is interchanged. The unperturbed system is now
\begin{equation}
   H_1 = \frac{{\bf p}^2}{2m}+V_1({\bf p},{\bf q})
\end{equation}
while initial conditions are kept. We note that momenta, scaling with $\sqrt{\lambda}$, force a dependence on $\lambda$ in the dual Hamiltonian but this dependence is harmless as a dual Lindstedt series should be considered with all leading order momenta going like $\sqrt{\lambda}{\bf p}$ fixing in this way the only dependence on $\lambda$ in the leading Hamiltonian. So, this has no effect in the proof of the dual KAM theorem.

Now, we will show that this interchange, obtained by rescaling, corresponds to a dual Lindstedt series. Hamilton equations are
\begin{eqnarray}
    \dot q_i(t)&=&\frac{p_i}{m}+\frac{\partial V_0}{\partial p_i}
    +\lambda\frac{\partial V_1}{\partial p_i} \\ \nonumber
    \dot p_i(t)&=&-\frac{\partial V_0}{\partial q_i}-\lambda\frac{\partial V_1}{\partial q_i}.
\end{eqnarray}
and let us rescale momenta as said above. One has
\begin{eqnarray}
    \dot q_i(t)&=&\sqrt{\lambda}\frac{p_i}{m}
    +\frac{1}{\sqrt{\lambda}}\frac{\partial V_0}{\partial p_i}
    +\sqrt{\lambda}\frac{\partial V_1}{\partial p_i} \\ \nonumber
    \sqrt{\lambda}\dot p_i(t)&=&-\frac{\partial V_0}{\partial q_i}
    -\lambda\frac{\partial V_1}{\partial q_i}
\end{eqnarray}
and this set of equations is consistent if we rescale time as $t\rightarrow\sqrt{\lambda}t$ giving
\begin{eqnarray}
\label{eq:dset}
    \dot q_i(t)&=&\frac{p_i}{m}   
    +\frac{\partial V_1}{\partial p_i} 
    +\frac{1}{\lambda}\frac{\partial V_0}{\partial p_i}\\ \nonumber
    \dot p_i(t)&=&
    -\frac{\partial V_1}{\partial q_i}
    -\frac{1}{\lambda}\frac{\partial V_0}{\partial q_i}
\end{eqnarray}
that are the equation one would obtain after the rescaling in (\ref{eq:res}). The rescaling in time so far introduced is consistent with the rescaling of the Hamiltonian. This approach is needed to obtain dual series as shown in \cite{fra1}. This proves that a dual Lindstedt series indeed exists that solves the set of equations (\ref{eq:dset}). Finally, we can write the dual Lindstedt series as
\begin{eqnarray}
    q_i(\sqrt{\lambda} t)&=&\sum_{n=0}^\infty \lambda^{-n}
    \hat q_i^n(\sqrt{\lambda} t) \\ \nonumber
    p_i(\sqrt{\lambda} t)&=&\lambda^\frac{1}{2}\sum_{n=0}^\infty \lambda^{-n}
    \hat p_i^n(\sqrt{\lambda} t)
\end{eqnarray}
recovering from all the rescaling. This produces $\alpha=0$ and $\beta=1/2$ in eqs. (\ref{eq:dlin}). We note as, now, the development parameter is $1/\lambda$ making meaningful the series in the limit $\lambda\rightarrow\infty$. We have seen the concept of duality in perturbation theory fully applied producing a dual series after interchanging the perturbation terms. We note that, after the application of the duality principle, what one has is again a standard perturbation problem sharing in this way all the properties of a small perturbation series. This implies that secular terms and small divisor problems are still there and we have to face them to extract meaningful results.

\section{KAM theorem and duality principle\label{kam}}

Behavior of integrable Hamiltonian systems under the effect of a small perturbation is ruled by KAM theorem. We have showed that a dual perturbation series for large perturbations does exist and so we ask if a similar result exists in this case. From the discussion above it is not difficult to see that answer to this question is affirmative. Our aim is to give a proof of a dual KAM theorem.

So, let us consider again a Hamiltonian system as
\begin{equation}
\label{eq:idsys}
   H=\frac{{\bf p}^2}{2m}+V_0({\bf p},{\bf q})+\lambda V_1({\bf p},{\bf q}).
\end{equation}
If $H_0=\frac{{\bf p}^2}{2m}+V_0({\bf p},{\bf q})$ is an integrable system, we can find a set of action-angle variables $\bf J$ and $\boldsymbol{\theta}$ in such away to have
\begin{equation}
   H=H_0(\mathbf{J})+\lambda H_1(\mathbf{J},\boldsymbol{\theta}).
\end{equation}
When $\lambda$ is a small parameter, the following theorem does hold \cite{arn1}

{\bf KAM Theorem}:{\sl If an unperturbed system is nondegenerate, then for sufficiently 
small conservative Hamiltonian perturbations, most non-resonant invariant 
tori do not vanish, but are only slightly deformed, so that in the phase space 
of the perturbed system, too, there are invariant tori densely filled with phase 
curves winding around them conditionally-periodically, with a number of 
independent frequencies equal to the number of degrees of freedom. 
These invariant tori form a majority in the sense that the measure of the 
complement of their union is small when the perturbation is small. 
}

Proof of this theorem was given in a series of classical papers by Kolmogorov, Arnol'd and Moser \cite{kolm,arn2,arn3,mos1,mos2}. This proof implies the convergence of the Lindstedt series notwithstanding the problem of small divisors and this was proved later by Eliasson \cite{elia}. So, turning back to our Hamiltonian (\ref{eq:idsys}) we now consider the limit $\lambda\rightarrow\infty$. By applying the duality principle to it, rescaling Hamiltonian and momenta, one has
\begin{equation}
   H=\frac{{\bf p}^2}{2m}+V_1({\bf p},{\bf q})+\frac{1}{\lambda} V_0({\bf p},{\bf q}).
\end{equation}
If $H_0=\frac{{\bf p}^2}{2m}+V_1({\bf p},{\bf q})$ is an integrable and non-degenerate system one can find a set of action-angle variables so that one can write
\begin{equation}
   H'=H'_1(\mathbf{J})+\frac{1}{\lambda} H'_0(\mathbf{J},\boldsymbol{\theta}).
\end{equation}
and for this Hamiltonian KAM theorem holds. So, the main conclusions drawn from KAM theorem for small perturbations on integrable systems holds also in the dual case for a class of strongly perturbed systems that admit again an integrable system at the leading order. This important dual KAM theorem will be applied in the next section. 

We give here a statement of the dual KAM theorem proved so far:

{\bf Dual KAM Theorem}:{\sl For sufficiently 
large conservative Hamiltonian perturbations, 
if a nondegenerate driving system exists, then its
most non-resonant invariant tori do not vanish, but are only slightly deformed, so that in the phase space of the perturbed system there are invariant tori densely filled with phase 
curves winding around them conditionally-periodically, with a number of 
independent frequencies equal to the number of degrees of freedom. 
These invariant tori form a majority in the sense that the measure of the 
complement of their union is small when the perturbation is large. 
}

This theorem says to us that, for this class of Hamiltonian systems, there is a finite region in parameter space with a fully developed chaos and that invariant tori should re-emerge if the perturbation is taken large enough. This phenomenon can be named {\sl tori reforming}. These tori are those of the driving system and do not coincide with those of the unperturbed system.

\section{Application\label{app}}

In this section we analyze a physical system that can display KAM behavior when there is a not so large perturbation. 
This clearly shows the effect of tori reforming when the perturbation becomes increasingly large.
A typical problem that is met in physics of plasma confinement and wherever a magnetic field and a plane wave are involved is described by the following Hamiltonian
\begin{equation}
   H=\frac{p^2}{2}+\frac{1}{2}\omega_0^2q^2-\lambda\alpha\cos(q-\omega t).
\end{equation}
The behavior of this system in the limit of small amplitude waves, $\lambda\rightarrow\infty$, is well known. It displays chaotic behavior when a given threshold is overcome \cite{kar1,kar2}. We can apply the dual KAM theorem to this system and show that it becomes non-chaotic in the limit $\lambda\rightarrow\infty$. Indeed, the application of the duality principle through variable rescaling gives to us
\begin{equation}
   H'=\frac{p^2}{2}-\alpha\cos(q-\omega t)+\frac{1}{2\lambda}\omega_0^2q^2.
\end{equation}
We see that the unperturbed Hamiltonian takes the form
\begin{equation}
   H_0=\frac{{p^0}^2}{2}-\alpha\cos(q^0-\omega \tau/\sqrt{\lambda})
\end{equation}
being $\tau=\sqrt{\lambda}t$. This, except for a spatial translation by $t$ that can be accomplished by a canonical transformation, is an integrable system. This can be straightforwardly realized noticing that such a canonical transformation gives back the pendulum Hamiltonian. As showed in \cite{arn1}, KAM theorem can be applied to such systems depending explicitly on time and, for this particular system, has been already proved in \cite{kar1,kar2}. So, dual KAM theorem is seen here to hold. This means that there exists a finite region in the parameter space where the system displays chaos or, stated otherwise, there exists a threshold that, when overcome, changes the motion from chaotic to a regular one.

As for the Lindstedt series, the leading order solution is given by
\begin{equation}
   q^0(\tau)=\frac{\omega}{\sqrt{\lambda}}\tau+
   2{\rm am}\left(\frac{1}{2}\sqrt{2\alpha(1+A)+\omega^2}\tau+\phi,
   \sqrt{\frac{4\alpha}{2\alpha(1+A)+\omega^2}}\right)
\end{equation}
being am the Jacobi amplitude such that $\sin({\rm am})={\rm sn}$ and $\cos({\rm am})={\rm cn}$ and $A,\ \phi$ two arbitrary constants fixed by initial conditions. In order to calculate the next-to-leading order correction, the following equation must be solved
\begin{equation}
   \ddot q^1(\tau)+\alpha\cos\left(q_0(\tau)-\omega \tau/\sqrt{\lambda}\right)q^1(\tau)
   =-\omega_0^2q^0(\tau)
\end{equation}
that becomes
\begin{eqnarray}
   \ddot q^1(\tau)&+&\alpha\left[1-2{\rm sn}^2
   \left(\frac{1}{2}\sqrt{2\alpha(1+A)+\omega^2}\tau+\phi,  
   \sqrt{\frac{4\alpha}{2\alpha(1+A)+\omega^2}}\right)\right]q^1(\tau)\\ \nonumber
   &=&-\omega_0^2q^0(\tau).
\end{eqnarray}
A possible condition to get this equation solved is given by \cite{fra9}
\begin{equation}
   \frac{4\alpha}{2\alpha(1+A)+\omega^2}\ll 1.
\end{equation}

\begin{figure}
\begin{center}
\subfigure[Poincar\'e section at $\alpha=0.5$, $\omega_0=1$ and $\omega=2$. Motion is regular.]{
\includegraphics[angle=-90, width=.45\textwidth]{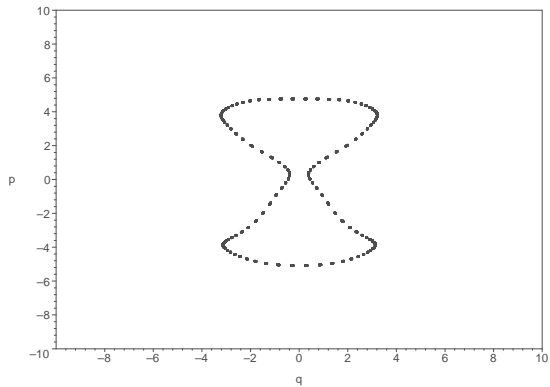}
}
\subfigure[Poincar\'e section at $\alpha=10$, $\omega_0=1$ and $\omega=2$. Tori are destroyed and fully developed chaos is seen.]{
\includegraphics[angle=-90, width=.45\textwidth]{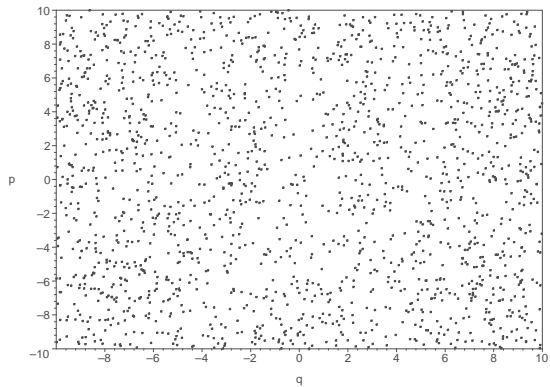}
}
\subfigure[Poincar\'e section at $\alpha=150$, $\omega_0=1$ and $\omega=2$. Tori reappear with some chaotic zones.]{
\includegraphics[angle=-90, width=.45\textwidth]{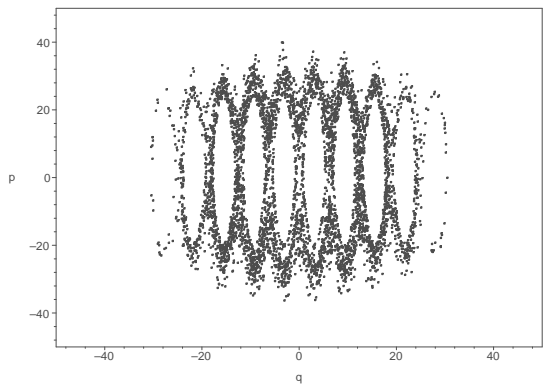}
}
\subfigure[Poincar\'e section at $\alpha=500$, $\omega_0=1$ and $\omega=2$. Tori are reformed.]{
\includegraphics[angle=-90, width=.45\textwidth]{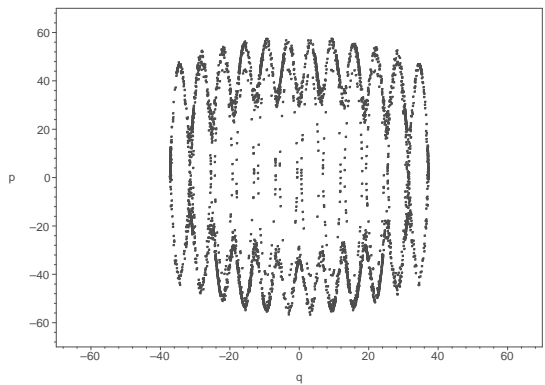}
}
\caption{\label{fig:fig1} Forced harmonic oscillator.}
\end{center}
\end{figure}

In order to see dual KAM theorem at work, we give some numerical evidence for this particular model: fig.\ref{fig:fig1} displays a sequence of Poincar\'e sections as perturbation increases reaching high values. A Poincar\'e section is a surface section of the phase space giving an immediate evidence of the behavior of the system. We can see the case of fully developed chaos from weak perturbation and a regular behavior back for a very strong perturbation clearly showing tori reforming. Reformed tori are those of a drifting pendulum giving helicoidal curves.

%


\section{Conclusions\label{cn}}

We have shown that a dual Lindstedt series can be obtained for a given Hamiltonian system. Correspondingly, a dual KAM theorem holds that produces tori reforming when a large perturbation is applied for a wide class of classical models. The approach is quite general and can also be extended to dissipative systems. But we would like to emphasize that for a large class of non-linear systems is open up the opportunity for their study in a vast range of parameter space.




\end{document}